\documentclass[a4paper]{lipics-v2021}

\usepackage{graphicx}

\usepackage{amsmath}
\usepackage{amsmath,bm}
\usepackage{amssymb}
\usepackage{adjustbox}

\usepackage{listings}
\usepackage{algorithm}
\usepackage{algpseudocode}

\usepackage{threeparttable}
\usepackage{multirow}
\usepackage{array}
\usepackage{tabularx}

\usepackage{todonotes}
\usepackage{color}
\usepackage{xcolor}

\usepackage{url}

\usepackage[shortlabels]{enumitem}
\setlist[enumerate]{nosep}

\title{Real-Time Guarantees for Critical Traffic in IEEE 802.1Qbv TSN Networks with Unscheduled and Unsynchronized End-Systems}

\titlerunning{} 

\author{Mohammadreza Barzegaran}{Technical University of Denmark Kongens Lyngby, Denmark}{mohba@dtu.dk}{}{}

\author{Niklas Reusch}{Technical University of Denmark Kongens Lyngby, Denmark}{nikre@dtu.dk}{}{}

\author{Luxi Zhao}{Technical University of Munich, Munich, Germany}{luxi.zhao@tum.de}{}{}

\author{Silviu S. Craciunas}{TTTech Computertechnik AG, Vienna, Austria}{silviu.craciunas@tttech.com}{}{}

\author{Paul Pop}{Technical University of Denmark Kongens Lyngby, Denmark}{paupo@dtu.dk}{}{}

\authorrunning{M. Barzegaran et. al.} 

\Copyright{Mohammadreza Barzegaran, Niklas Reusch, Luxi Zhao, Silviu S. Craciunas, and Paul Pop} 



\category{} 

\relatedversion{} 
\hideLIPIcs  

\EventEditors{}
\EventNoEds{2}
\EventLongTitle{}
\EventShortTitle{}
\EventAcronym{}
\EventYear{}
\EventDate{}
\EventLocation{}
\EventLogo{}
\SeriesVolume{}
\ArticleNo{}

\begin{document}
\nolinenumbers

\maketitle

\begin{abstract}
Time-Sensitive Networking (TSN) aims to extend the IEEE 802.1Q Ethernet standard with real-time and time-aware capabilities. Each device's transmission of time-critical frames is done according to a so-called Gate Control List (GCL) schedule via the timed-gate mechanism described in IEEE 802.1Qbv. Most schedule generation mechanisms for TSN have a constraining assumption that both switches and end-systems in the network must have at least the TSN capabilities related to scheduled gates and time synchronization. However, many TSN networks use off-the-shelf end-systems, e.g., for providing sensor data, which are not scheduled and/or synchronized.

In this paper, we propose a more flexible scheduling strategy that considers a worst-case delay analysis within the scheduling synthesis step, leveraging the solution's optimality to support TSN networks with unscheduled and unsynchronized end-systems while still being able to guarantee bounded latency for critical messages. Our method enables real-world systems that feature off-the-shelf microcontrollers and sensor nodes without TSN capabilities connected to state-of-the-art TSN networks to communicate critical messages in a real-time fashion. We evaluate our approach using both synthetic and real-world test cases, comparing it with existing scheduling mechanisms. Furthermore, we use OMNET++ to validate the generated GCL schedules.
\end{abstract}

\section{Introduction}

Standardized communication protocols allowing safety-critical communication with real-time guarantees are becoming increasingly relevant in application domains beyond aerospace, e.g., in industrial automation and automotive. Time-Sensitive Networking (TSN)~\cite{8021tsn} amends the standard Ethernet protocol with real-time capabilities ranging from clock synchronization and frame preemption to redundancy management and schedule-based traffic shaping~\cite{Craciunas16:RTNS}. These novel mechanisms allow standard best-effort (BE) Ethernet traffic to coexist with isolated and guaranteed scheduled traffic (ST) within the same multi-hop switched Ethernet network. The main enablers of this coexistence are a network-wide clock synchronization protocol (802.1ASrev~\cite{IEEEASrev}) defining a global network time, known and bounded device latencies (e.g., switch forwarding delays), and a Time-Aware Shaper (TAS) mechanism~\cite{802.1Qbv} with a global communication schedule implemented in so-called Gate Control Lists (GCLs), facilitating ST traffic with bounded latency and jitter in isolation from BE communication. The TAS mechanism is implemented as a gate for each transmission queue that either allows or denies the sending of frames according to the configured GCL schedule.

The schedule generation for GCLs~\cite{Craciunas16:RTNS, Oliver18, Pop16} has focused on enforcing deterministic transmission with isolation and compositional system design for ST flows requiring end-to-end guarantees. Craciunas et al.~\cite{Craciunas16:RTNS} proposed a Satisfiability Modulo Theories (SMT)-based method that binds a transmission slot (window) to at most one ST frame producing a schedule with 0-jitter and defined end-to-end latencies critical flows. Deterministic transmission and forwarding is enabled by isolating critical streams either in time or in different traffic classes. A drawback of this model is that the solution space may be significantly reduced, and the number of GCL entries may be too high for the storage capabilities of TSN devices~\cite{ReuschWFCS20}. In~\cite{Oliver18}, multiple ST frames are allowed to share a transmission slot (window), and hence they may delay each other (if they are not isolated in the space domain). Hence,~\cite{Oliver18} relaxes the 0-jitter assumption while still maintaining deterministic end-to-end temporal behavior. However, both approaches necessitate that the traffic leaves the sending end systems in a scheduled and synchronized fashion (i.e., requiring TSN capabilities on both end systems and switches) since the schedule controls the transmission and forwarding of frames from sending node to receiving node within the network, requiring that the interference between ST frames is either 0 or bounded by the schedule construction. 

The limitation of previous approaches requiring end systems to be scheduled and synchronized to the network, i.e., to have 802.1Qbv and 802.1ASrev TSN capabilities, is often not realistic since many systems use, e.g., off-the-shelf microcontrollers, industrial PCs, and sensor nodes without TSN mechanisms. Hence, there is a need for GCL synthesis mechanisms that can schedule TSN networks with unscheduled and unsynchronized end systems in real-world applications. Other work, c.f.~\cite{ReuschWFCS20, Hellmanns20}, introduce class-based GCLs without the requirement for synchronized end systems. However, their scheduling models are quite limited, requiring all sending windows to be aligned among all switches and not using formal verification methods for the delay analysis within the scheduling step.

In this paper, we relax the requirement that end-systems need to be synchronized and/or scheduled and consider relative offsets of windows on different nodes. We intertwine the worst-case delay analysis from~\cite{Zhao20:JIoT} with the scheduling step in order to generate correct schedules where the end-to-end requirements of ST flows are met. Furthermore, we describe different TSN scheduling approaches that have been proposed in the literature (see Table~\ref{tab:problemcompositions} for an overview) and compare them to our flexible window-based approach. We define the analysis-driven window optimization problem resulting from our more flexible approach with the goal to be able to enlarge the solution space, reduce computational complexity, and apply it to end-systems without TSN mechanisms. Depending on industrial applications' requirements, our evaluation can help system designers choose the most appropriate combination of configurations for their use-case. The main contributions of the paper are:
\begin{enumerate}
	\item[$\bullet$] We propose a novel flexible window-based scheduling method that does not individually schedule ST frames and flows but rather schedules open gate windows for individually scheduled queues.
	Hence, we can support non-deterministic queue states and hence networks with unscheduled and/or unsynchronized end-systems by integrating the WCD Network Calculus (NC) analysis into the scheduling step. The NC analysis is used to construct a worst-case scenario for each flow to check its schedulability, considering arbitrary arrival times of flows and the open GCL window placements.
	\item[$\bullet$] We compare and evaluate our flexible window-based scheduling method with existing scheduling methods for TSN networks. To the best of our knowledge, this is the first work comparing such mechanisms for TSN.
	\item[$\bullet$] We propose a proxy function as an extension for the analysis in~\cite{Zhao20:JIoT} and implement it in our problem formulation.
	\item[$\bullet$] We formulate and solve a window optimization problem that uses the proxy function and provides timing guarantees for real-time flows even in systems with unscheduled and unsynchronized end systems.
	\item[$\bullet$] We evaluate the proposed approach on both synthetic and real-world test cases, validating the correctness and the scalability of our implementation and use the OMNET++ simulator to validate the generated solutions.
\end{enumerate}

We start by presenting a review of related research, focusing on the existing scheduling mechanisms that we compare our work to, in Sect.~\ref{sec:relwork}. We introduce the system, network, and application models, as well as a description of the main TSN standards, in Sect.~\ref{sec:model}. We outline the problem formulation in Sect.~\ref{sec:ProFormu}. In Sect.~\ref{sec:solution}, we present the novel scheduling mechanism and the optimization strategy based on the Constraint Programming (CP) followed by the comparison and evaluation results in Sect.~\ref{sec:Evaluation}. We conclude the paper in Sect.~\ref{sec:Conclusion}.	

\section{Related Work}
\label{sec:relwork}
Since its introduction, significant research efforts have focused on studying the real-time properties of Time-Sensitive Networking enabled by mechanisms such as 802.1Qbv and 802.1Asrev and the resulting TSN scheduling problems. In~\cite{Frank16} the TSN scheduling problem is reduced to having one single queue for ST traffic and solving it using Tabu Search, as well as an optimization that reduces the number of guard-bands in order to optimize bandwidth usage. The work in~\cite{Hellmans20} proposes a scheduling model for TSN networks in industrial automation with different traffic types and a hierarchical scheduling procedure for isochronous traffic.

The most relevant work for providing real-time communication properties in TSN networks (and to which we compare our approach to) has been presented in~\cite{Craciunas16:RTNS, Oliver18, Pop16}.
The main goal of these works is to allow temporal isolation and compositional system design for ST flows with end-to-end guarantees and deterministic communication behavior. The work in~\cite{Craciunas16:RTNS}, which we call \emph{0GCL}, presents a solution that generates the most deterministic and strict schedules for TSN. The scheduling constraints result in frame transmission with $0$ jitter in each hop and guaranteed end-to-end latency. The deterministic behavior comes from the frame-based scheduling approach, which does not allow any variation in the transmission times and thus no interference between ST frames even when placed in the same queue. The work in~\cite{Pop16} maintains the strict determinism but solves the problem using heuristics instead of SMT-solvers in order to improve scalability. Additionally, the ST schedules are generated such that the worst-case delays of AVB streams are minimized. The scheduling approach in~\cite{Oliver18}, which we call \emph{Frame-to-Window-based}, relaxes the 0-jitter constraint of~\cite{Craciunas16:RTNS} and thus allows more variance in the transmission times of frames along the hops of the TSN network. This increases the solution space at the expense of increased complexity in the correctness constraints. This method is also window-based, as is ours, but requires a unique mapping between GCL windows and frames in order to avoid non-deterministic delays in the queues.

The mentioned approaches require an essential property, namely, that both the end-systems and switches have TSN capabilities (i.e., are in a scheduled and synchronized fashion). Thus, the schedule controls the transmission and forwarding of frames from a sending node to a receiving node within the network, requiring that the interference between ST frames is either 0 or bounded by the schedule construction. The open gate windows are then either a result of the frame transmission schedule~\cite{Craciunas16:RTNS, Pop16} or are uniquely associated with predefined subsets of frames~\cite{Oliver18}. However, the above property is a significant limitation. In many use cases, the endpoints (e.g., off-the-shelf) sensors, microcontrollers, industrial PCs, and edge devices do not have TSN capabilities (i.e., the 802.1Qbv timed-gate mechanism or the time-synchronization protocol of 802.1ASrev). The different scheduling mechanisms described above that we compare our work to are summarized in Table~\ref{tab:problemcompositions}.

The work in~\cite{ReuschWFCS20} proposed a similar window-based approach (WND) but without the CP-based solution and with a very limited model since the offsets of the windows on different nodes are not considered, thereby essentially limiting the mechanisms by requiring all windows to be aligned among all switches. Moreover,~\cite{ReuschWFCS20} uses a less advanced analysis step (c.f.~\cite{Zhao18:Access}) in the scheduling decisions and a simpler heuristic approach. In~\cite{Hellmanns20}, a class-based GCL was proposed without per-flow scheduling nor requirement of synchronized end-systems. The method proposed in~\cite{Hellmanns20} for calculating the worst-case delay bounds, unlike our approach, is not based on any formal verification methods (e.g., network calculus). The worst-case delay calculation is overly pessimistic for small-scale networks and not safe for large-scale networks. 

Classical approaches like strict priority (SP) and AVB~\cite{802.1BA} do not require a time-gate mechanism and also work with unscheduled end-systems. In order to provide response-time guarantees, a worst-case end-to-end timing analysis through methods like
network calculus~\cite{Schmitt03, DeAzua14} or Compositional Performance Analysis (CPA)~\cite{Diemer12} are used. In~\cite{Zhao17, Zhao14, Boyer16}, the rate-constrained (RC) flows of TTEthernet~\cite{sae-as6802, steiner11:CRC} are analyzed using network calculus. Other works, such as~\cite{Wanderler06a, Khanh14}, study the response-time analysis for TDMA-based networks under the strict priority (SP) and weighted round-robin (WRR) queuing policies. Zhao et al.~\cite{Zhao20:JIoT} present a worst-case delay analysis, which we use in this paper, for determining the interference delay between ST traffic on the level of flexible GCL windows.

In~\cite{9157984}, the authors present hardware enhancements to standard IEEE 802.1Qbv bridges (along with correctness constraints for the schedule generation) that remove the need for the isolation constraints between frames scheduled in the same egress queue defined in~\cite{Craciunas16:RTNS}. Another hardware adaptation for TSN bridges, which has been proposed by Heilmann et al.~\cite{10.1145/3314206.3314207} is to increase the number of non-critical queues in order to improve the bandwidth utilization without impacting the guarantees for critical messages.

\section{System Model}
\label{sec:model}
This section defines our system model for which we summarize the notation in Table.~\ref{tab:notation}.
	
\begin{table}[t]
\caption{Summary of notations}
\label{tab:notation}
\setlength{\tabcolsep}{3pt}
\begin{tabular}{|p{120pt}|p{245pt}|}
\hline
Symbol & System model\\
\hline
$G=(\bm{V},\bm{E})$ & Network graph with the set of nodes ($\bm{V}$) and links ($\bm{E}$)\\
$[v_a, v_b] \in \bm{E}$&Link\\
$[v_a, v_b].C, [v_a, v_b].mt$&Link speed and macrotick\\
$p \in P$&Output port\\
$p.Q$&Eight priority queues in an output port $p$\\
$q\in p.Q_{ST}$&A queue used for ST traffic in $p$\\
$\left<\phi, w, T\right>_q$&GCL configuration for a queue $q\in p.Q_{ST}$, where $q.\phi$, $q.w$, and $q.T$ are the window offset, the window length , and the window period for a queue $q$, respectively.\\
$f.l, f.T, f.P, f.D $&Payload size, period, priority, and deadline of a flow $f\in \mathcal{F}$\\
$f.r=[[v_1,v_2], ..., [v_{n-1},v_n]]$&Route for a flow $f\in \mathcal{F}$\\
\hline
\end{tabular}
\label{tab1}
\end{table}

\subsection{Network Model}
We represent the network as a directed graph $G=(\bm{V},\bm{E})$ where $\bm{V}=\bm{ES}\bigcup \bm{SW}$ is the set of end systems (ES) and switches (SW) (also called nodes), and $\bm{E}$ is the set of bi-directional full-duplex physical links. An ES can receive and send network traffic while SWs are forwarding nodes through which the traffic is routed. 
The edges $\bm{E}$ of the graph represent the full-duplex physical links between two nodes, $\bm{E}\subseteq \bm{V}\times \bm{V}$. If there is a physical link between two nodes $v_a, v_b\in \bm{V}$, then there exist two ordered tuples $[v_a, v_b], [v_b,v_a]\in \bm{E}$. An equivalence between output ports $p\in P$ and links $[v_a,v_b]\in \bm{E}$ can be drawn as each output port is connected to exactly one link. 
A link $[v_a, v_b]\in \bm{E}$ is defined by the link speed $C$ (Mbps), propagation delay $d_p$ (which is a function of the physical medium and the link length), and the macrotick $mt$. The macrotick is the length of a discrete time unit in the network, defining the granularity of the scheduling timeline~\cite{Craciunas16:RTNS}. Without loss of generality, we assume $d_p=0$ in this paper.
	
	As opposed to previous work, we do not require that end-system are either synchronized or scheduled. Since ESs can be unsynchronized and unscheduled, they transmit frames according to a strict priority (SP) mechanism. Switches still need to be synchronized and scheduled using the 802.1ASrev and 802.1Qbv, respectively. 

\subsection{Switch Model}
\label{sec:sw-model}
\begin{figure}
     \centering
     \begin{subfigure}[b]{0.45\textwidth}
         \centering
         \includegraphics[width=\textwidth]{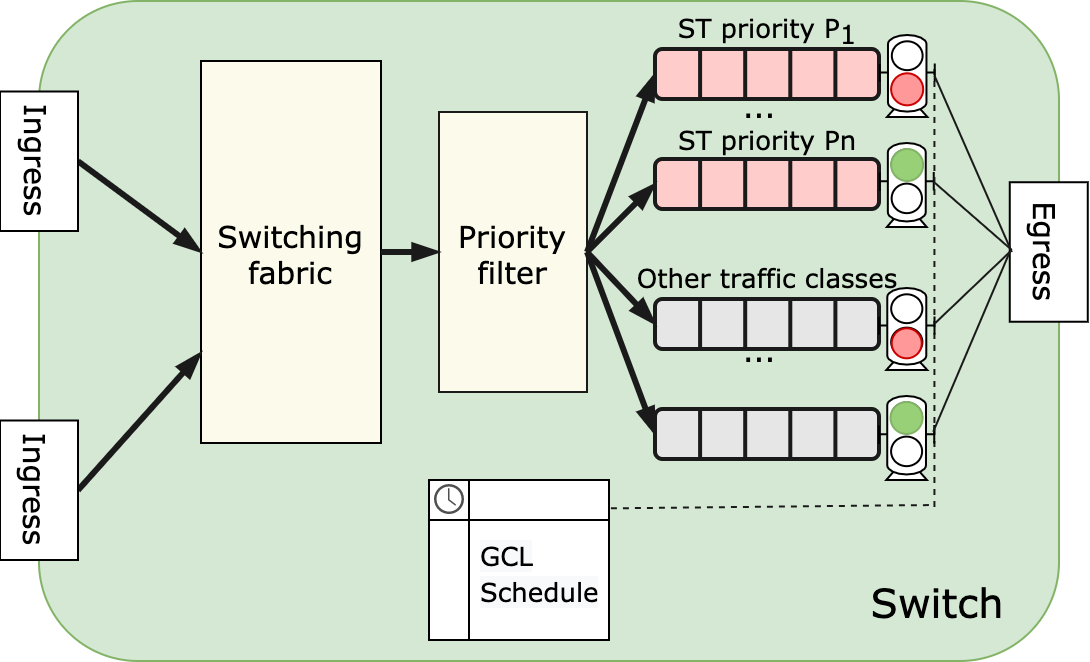}
         \caption{TSN Switch Internals}
         \label{fig:TSNswitch}
     \end{subfigure}
     \hfill
     \begin{subfigure}[b]{0.38\textwidth}
         \centering
         \includegraphics[width=\textwidth]{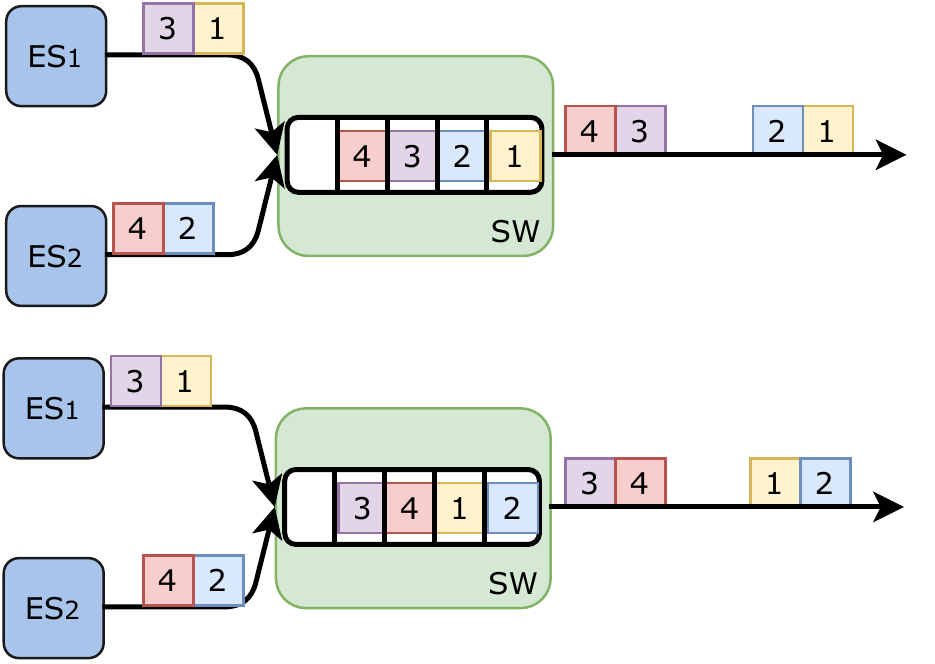}
         \caption{Queue states (inspired by~\cite{9157984}).}
         \label{fig:non_det}
     \end{subfigure}
     \caption{Switch internals and non-deterministic queue states due to variations in frame arrivals from unscheduled and/or unsynchronized ESs.}
\end{figure}

Fig.~\ref{fig:TSNswitch} depicts the internals of a TSN switch. The switching fabric decides, based on the internal routing table to which output port $p$ a received frame will be forwarded. Each egress port has a priority filter that determines in which of the available $8$ queues (traffic-classes) of that port a frame will be put. Each priority queue $q\in p.Q$ stores the frames to be forwarded in first-in-first-out (FIFO) order. Similar to~\cite{Craciunas16:RTNS}, a subset of the queues ($p.Q_{ST}$) is used for ST traffic, while the rest ($p.\overline{Q}$) are used for non-critical communication. As opposed to regular 802.1Q bridges, where enqueued frames are sent out according to their respective priority, in 802.1Qbv bridges, there is a Time-Aware Shaper (TAS), also called timed-gate, associated with each queue and positioned behind it. A timed-gate can be either in an \emph{open} (\emph{O}) or \emph{closed} (\emph{C}) state. When the gate is open, traffic from the respective queue is allowed to be transmitted, while a closed gate will not allow transmission, even if the queue is not empty. When multiple gates are open simultaneously, the highest priority queue has preference, blocking others until it is empty or the corresponding gate is closed. The 802.1Qbv standard includes a mechanism to ensure that no frames can be transmitted beyond the respective gate's closing point. This look-ahead checks whether the entire frame present in the queue can be fully transmitted before the gate closes and, if not, it will not start the transmission.

The state of the queues is encoded in a GCL. A fundamental difference to the similar technology TTEthernet (SAE AS6802)~\cite{sae-as6802}, is that in TSN the scheduling is done on the level of traffic-classes instead of on an individual frame level~\cite{CraciunasETR17}. Hence, an imperfect time synchronization, frame loss, ingress policing (c.f.~\cite{Craciunas16:RTNS}), or the variance in the arrival of frames from unscheduled and/or unsynchronized ESs may lead to non-determinism in the state of the egress queues and, as a consequence, in the whole TSN network. If the state of the queue is not deterministic at runtime, the order and timing of the sending of ST frames can vary dynamically. In Fig.~\ref{fig:non_det}, the schedule for the queue of the (simplified) switch SW, opens for two frames and then, sometime later, for the duration of another two frames. The arrival of frames from unscheduled and/or unsynchronized end systems may lead to a different pattern in the egress queue of the switch, as illustrated in the top and bottom figures of Fig.~\ref{fig:non_det}. Note that we do not actually know the arrival times of the frames, and what we depict in the figure are just two scenarios to illustrate the non-determinism. There may be scenarios where one of the frames, e.g., frame ``2'', arrives much later. This variance makes it impossible to isolate frames in windows and obtain deterministic queue states, and, as a consequence, deterministic egress transmission patterns, as required by previous methods for TSN scheduling (e.g.~\cite{Craciunas16:RTNS, Oliver18, Pop16, Frank16}). We refer the reader to~\cite{Craciunas16:RTNS} for an in-depth explanation of the TSN non-determinism problem. 

The queue configuration is expressed by a tuple $q=\langle Q_{ST}, \overline{Q} \rangle$. The decision in which queue to place frames is taken either according to the priority code point (PCP) of the VLAN tag or according to the priority assignment of the IEEE~802.1Qci mechanism. In order to formulate the scheduling problem, the GCL configuration is defined as a tuple $\left<\phi, w, T\right>_q$ for each queue $q \in p.Q_{ST}$ in an output port $p$, with the window offset $\phi$, window length $w$ and window period $T$.

\subsection{Application Model}
\label{sec:app-model}
The traffic class we focus on in this paper is scheduled traffic (ST), also called time-sensitive traffic. ST traffic is defined as having requirements on the bounded end-to-end latency and/or minimal jitter~\cite{Craciunas16:RTNS}. Communication requirements of ST traffic itself are modeled with the concept of flows (also called streams), representing a communication from one sender (talker) to one or multiple receivers (listeners). We define the set of ST flows in the network as $\mathcal{F}$. A flow $f\in \mathcal{F}$ is expressed as the tuple $\left<l, T, P, D\right>_f$, including the frame size, the flow period in the source ES, the priority of the flow, and the required deadline representing the upper bound on the end-to-end delay of the flow.
The route for each flow is statically defined as an ordered sequence of directed links,e.g., a flow $f\in F$ sending from a source ES $v_1$ to another destination ES $v_n$ has the route $r=[[v_1,v_2], ..., [v_{n-1},v_n]]$. Without loss of generality, the notation is simplified by limiting the number of destination ES to one, i.e., unicast communication. Please note that the model can be easily extended to multicast communication by adding each sender-receiver pair as a stand-alone flow with additional constraints between them on the common path.

\section{Problem Formulation}
\label{sec:ProFormu}

\begin{figure*}[!b]
	\centering
	\includegraphics[width=0.90\textwidth]{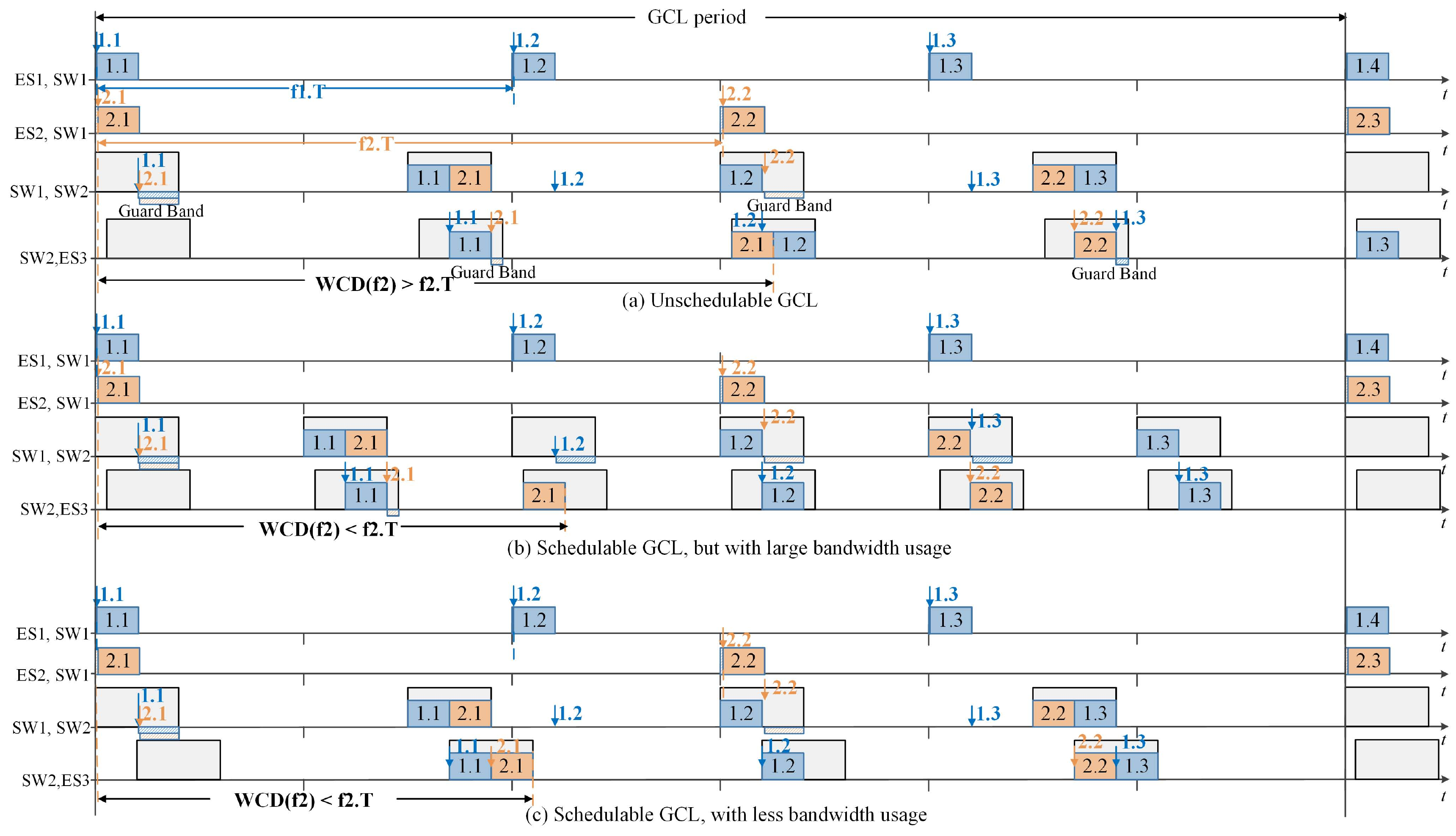}
	\caption{\label{fig:MotiExample} Motivational example showing the importance of optimizing the windows. Note that the arrival times of frames are not known beforehand, hence we decided to illustrate in each configuration (a) to (c) an arrival scenario that leads to the worst-case delay for frame $f_2$.}
\end{figure*}

The problem formulation is as follows:
Given (1) a set of flows~$\mathcal{F}$ with statically defined routes~$\mathcal{R}$, and (2) a network graph~$G$, we are interested in determining GCLs, which is equivalent to determining (i) the offset of windows $q.\phi$, (ii) the length of windows $q.w$, and (iii) the period of windows $q.T$ such that the deadlines of all flows are satisfied and the overall bandwidth utilization (defined in Sect.~\ref{sec:objectivefunction}) is minimized.

Let us illustrate the importance of determining optimized windows. Recall that with flexible window-based scheduling we do not know the arrival times of frames, and frames of different ST flows may interfere with each other. Frames that arrive earlier will delay frames that arrive later; also, a frame may need to wait until a gate is open, or arrive at a time just before a gate closure and cannot fit in the interval that remains for transmission. 

We illustrate in Fig.~\ref{fig:MotiExample} three window configurations (a), (b), and (c), motivating the need to optimize the windows. The vertical axis represents each egress port in the network, and the horizontal axis represents the timeline. The tall grey rectangles give the gate open time for a priority queue. As mentioned, we do not know the arrival times of the frames, thus it is necessary to provide a formal analysis method to ensure real-time performance. In this paper, the timing guarantees are provided for a given window configuration using the 
Network Calculus (NC)-based approach from~\cite{Zhao20:JIoT} to determine the worst-case end-to-end delay bounds (WCDs) for each flow. In the motivational example, the WCDs are determined by constructing a worst-case scenario for each flow. Hence, in Fig.~\ref{fig:MotiExample} we show worst-case scenarios. The red and blue rectangles in the figure represent ST frames' transmission. There are two periodic flows $f_1$ (blue rectangles) and $f_2$ (red rectangles) with the same frame size and priority. We use the arrows pointing down to mark the arrival time for ST frames creating the worst-case case for $f_2$. Let us assume that the deadline of each flow equals its period $f_i.T$. In each configuration in Fig.~\ref{fig:MotiExample} we show that arrival scenario which would lead to the worst-case situation for frame $f_2$, i.e., the largest WCD for $f_2$.

Since the ESs are unscheduled (without TAS) and/or without time synchronization capabilities, the frames can arrive and be transmitted by the ES at any time (the offset of a periodic flow on the ES is in an arbitrary relationship with the offsets of the windows in the SWs). However, on the switches, frame transmissions are allowed to be forwarded only during the scheduled window. The worst-case for $f_2$ happens when the frame (1.1) of $f_1$ arrives on [$ES_1, SW_1$] slightly earlier than the frame (2.1) of $f_2$ arrives on [$ES_2, SW_1$], and at the same time, they arrive on the subsequent egress port [$SW_1, SW_2$] at a time when the remaining time during the current window is smaller than the frame transmission time. In this case, the guard band delays the frames until the next window slot.

With the windows configuration in Fig.~\ref{fig:MotiExample}(a), the WCD of $f_2$ is larger than its deadline, i.e., $WCD(f_2)>f_2.T$, hence, $f_2$ is not schedulable. If the window period is narrowed down, as shown in Fig.~\ref{fig:MotiExample}(b), the WCD of $f_2$ satisfies its deadline. However, there is a large bandwidth usage occupied by the windows. Fig.~\ref{fig:MotiExample}(c) uses the same window period as in Fig.~\ref{fig:MotiExample}(a) but changes the window offset. As can be seen in the figure, the WCD of the flow $f_2$ is also smaller than its deadline $f_2.T$, and compared with Fig.~\ref{fig:MotiExample}(b), the bandwidth usage of the windows is reduced. With the increasing complexity of the network, e.g., multiple flows joining and leaving at any switch in the network and/or an increased number of ST flows and priorities, an optimized window configuration cannot be done manually; therefore, optimization algorithms are needed to solve this problem.

\section{Optimization Strategy}
\label{sec:solution}
The problem presented in the previous section is intractable. The decision problem associated with the scheduling problem has been proved to be (NP)-complete in the strong sense~\cite{sinnen2007task}. Hence, we use an optimization strategy, called Constraint Programming-based Window Optimization (CPWO), which is more suitable to find good solutions in a reasonable time.

\begin{figure}[!b]
	\centering
	\includegraphics[width=0.65\textwidth]{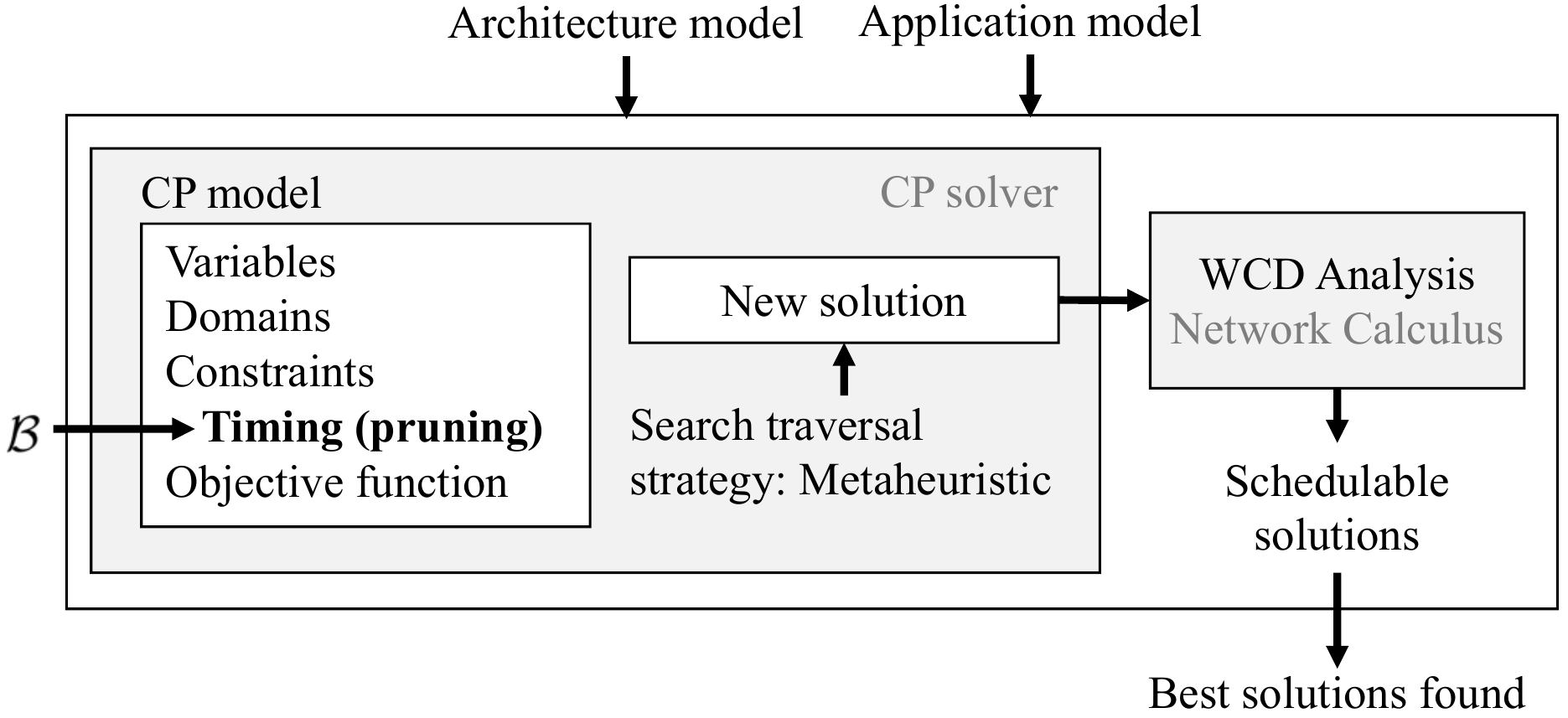}
	\caption{Overview of our proposed CPWO optimization strategy}
	\label{fig:strategy}
\end{figure}

CPWO takes as the inputs the architecture and application models and outputs a set of the best solutions found during search (see Fig.~\ref{fig:strategy}). We use Constraint Programming (CP) to search for solutions (the ``CP solver'' box). 
CP performs a systematic search to assign the values of variables to satisfy a set of constraints and optimize an objective function, see the ``CP model'' box: the sets of variables are defined in Sect.~\ref{sec:CPModel}, the constraints in Sect.~\ref{sec:constraints} and the objective function in Sect.~\ref{sec:objectivefunction}. A feasible solution is a valid solution that is schedulable, i.e., the worst-case delays (WCDs) of streams are within their deadlines. Since it is impractical to check for schedulability within a CP formulation, we employ instead the Network Calculus (NC)-based approach from~\cite{Zhao20:JIoT} to determine the WCDs, see the ``WCD Analysis'' box in Fig.~\ref{fig:strategy}. The WCD Analysis is called every time the CP solver finds a ``new solution'' which is valid with respect to the CP constraints. The ``new solution'' is not schedulable if the calculated latency upper bounds are larger than the deadlines of some critical streams.

Although CP can perform an exhaustive search and find the optimal solution, this is infeasible for large networks. Instead, CPWO employs two strategies to speed up the search to find optimized schedulable solutions in a reasonable time, at the expense of optimality. 

(i) A metaheuristic search traversal strategy: CP solvers can be configured with user-defined search strategies, which enforce a custom order for selecting variables for assignment and for selecting the values from the variable's domain. Here, we use a \emph{metaheuristic} strategy based on Tabu Search~\cite{burke2005search}.

(ii) A timing constraint specified in the CP model that prunes the search space: Ideally, the WCD Analysis would be called for each new solution. However, an NC-based analysis is time-consuming, and it would slow down the search considerably if called each time the CP solver visits a new valid solution. Hence, we have introduced ``search pruning'' constraints in the CP model (the ``Timing (pruning)'' constraints in the ``CP model'' box in Fig.~\ref{fig:strategy}), explained in detail in Sect.~\ref{sec:proxy}. 

These timing constraints implement a crude analysis that indicates if a solution may be schedulable and are solely used by the CP solver to eliminate solutions from the search space. These constraints may lead to both ``relaxed-pruning'' scenarios that are actually unschedulable or ``aggressive-pruning'' scenarios that eliminate solutions that are schedulable. The proxy function (pruning constraint) can thus be parameterized to trade-off runtime performance for search-space pruning in the CP-model.

The timing constraints assume that for a given stream, its frames in a queue will be delayed by other frames in the same queue, including a backlog of frames of the same stream. A parameter $\mathcal{B}$ is used to adjust the number of frames in the backlog, tuning the pruning level of the CP model's timing constraints. Note that NC still checks the actual schedulability, so it does not matter if the CP analysis is too relaxed---this will only prune fewer solutions, slowing down the search. However, using overly aggressive pruning runs the risk of eliminating schedulable solutions of good quality. We consider that $\mathcal{B}$ is given by the user, controlling how fast to explore the search space. In the experiments, we adjusted $\mathcal{B}$ based on the feedback from the WCD Analysis and the pruning constraint. If, during a CPWO run, the pruning constraint from Sect.~\ref{sec:proxy} was invoked too often, we decreased $\mathcal{B}$, as it was pruning too aggressively; otherwise, if the WCD analysis was invoked too often and was reporting that the solutions were schedulable, we increased $\mathcal{B}$.

We first define the terms needed for the CP model in Table.~\ref{tab:cpmodelterms}. Then, we continue with the definition of the objective function, model variables, and constraints of the CP model.

\begin{table}[b]
\small
\caption{Definition of terms used in CP model formulation}
\label{tab:cpmodelterms}
\centering
\small
\begin{tabular}{l|l}
\hline
\textbf{Term}&\textbf{Definition}\\
\hline
\hline
$\mathcal{N}(P)$  & Total number of windows assigned to priority queues\\
$\mathcal{K}(p)$  & Hyperperiod of the port~$p$\\
$\mathcal{L}(q)$  & Maximum size of any frame from all flows assigned to the queue~$q$\\
$\mathcal{GB}(q)$ & Maximum transmission time of ST frames competing in the queue $q$\\
$\mathcal{R}(q)$ & All flows assigned to the queue $q$\\
$\mathcal{X}(q)$ & All flows arriving from a switch and assigned to the queue $q$\\
\hline
\end{tabular}
\end{table}

\subsection{Objective Function}
\label{sec:objectivefunction}
The CP solver uses the objective function~$\Omega$, which minimizes the average bandwidth usage: 
\begin{align}
   \label{eq:avgbandwidth}
        &\forall p \in P,\forall q \in p.Q: \Omega = \frac{\sum \frac{q.w}{q.T}}{\mathcal{N}(P)}.
\end{align}

\noindent The average bandwidth usage is calculated as the sum of each window's utilization, i.e., the window length over its period, divided by the total number of windows in the CP model. Note that solutions found by a CP solver are guaranteed to satisfy the constraints defined in Sect.~\ref{sec:constraints}. In addition, the schedulability is checked with the NC-based WCD Analysis~\cite{Zhao20:JIoT}. 

\subsection{Variables}
\label{sec:CPModel}
The model variables are the offset, length, and period of each window, see Sect.~\ref{sec:sw-model}. For each variable, we define a domain which is a set of finite values that can be assigned to the variable.
CP decides the values of the variables as an integer from their domain in each visited solution during the search.
The domains of offset $q.\phi$, length $q.w$, and period $q.T$ variables are defined, respectively, by
\begin{equation}
   \label{eq:defdomains}
   \begin{split}
        &\forall p \in P,\forall q \in p.Q:\\
        &0 < q.T \leq \frac{\mathcal{K}(p)}{[v_a, v_b].mt},\quad 0 \leq q.\phi \leq \frac{\mathcal{K}(p)}{[v_a, v_b].mt},\\
        &\frac{\mathcal{L}(q)}{[v_a, v_b].mt \times [v_a, v_b].C} + \mathcal{GB}(q) \leq q.w \leq \frac{\mathcal{K}(p)}{[v_a, v_b].mt}.
   \end{split}
\end{equation}
The domain of the window period is defined in the range from 0 to the hyperperiod of the respective port~$p$, i.e., the Least Common Multiple (LCM) of all the flow periods forwarded via the port. The window period is an integer and cannot be zero. The domain of the window offset is defined in the range from 0 to the hyperperiod of the respective port~$p$. Finally, the domain of the window length is defined in the range from minimum accepted window length to the hyperperiod of the respective port~$p$. The minimum accepted window length is the length required to transfer the largest frame from all flows assigned to the queue~$q$, protected by the guard band $\mathcal{GB}(q)$ of the queue.
A port~$p$ is attached to only one link~$[v_a, v_b]$; and values and domains are scaled by the macrotick $mt$ of the respective link.

\subsection{Constraints}
\label{sec:constraints}
The first three constraints need to be satisfied by a valid solution: (1) the window is valid, (2) two windows in the same port do not overlap, and (3) windows' bandwidth is not exceeded. 
The last two constraints reduce the search space by restricting the periods of (4) queues and (5) windows to harmonic values in relation to the hyperperiod. Harmonicity may eliminate some feasible solutions but we use this heuristic strategy to speed up the search.

\noindent (1) The \textbf{Window Validity Constraint} (Eq.~\eqref{eq:windowproperty}) states that the offset plus the length of a window should be smaller or equal to the window's period:
\begin{equation}
   \label{eq:windowproperty}
   \begin{split}
        &\forall p \in P,\forall q \in p.Q: \quad (q.w + q.\phi) \leq q.T.
   \end{split}
\end{equation}

\noindent(2) \textbf{Non-overlapping Constraint} (Eq.~\eqref{eq:overlap}). Since we search for solutions in which windows of the same port do not overlap, the opening or closing of each window on the same port (defined by its offset and the sum of its offset and length, respectively) is not in the range of another window, over all period instances:
\begin{align}
   \label{eq:overlap}
        &\forall p \in P,\forall q \in p.Q, \forall q' \in p.Q, T_{q,q'}= max(q.T, q'T), \forall a \in [0,T_{q,q'}/q.T),\forall b \in [0,T_{q,q'}/q'.T):\nonumber\\
        & (q.\phi + q.w+ a \times q.T) \leq (q'.\phi + b \times q'.T) \vee (q'.\phi + q'.w + b \times q'.T) \leq (q.\phi+ a \times q.T)
\end{align}
\noindent (3) The \textbf{Bandwidth Constraint} (Eq. (\ref{eq:bandwidth})) ensures that all the windows have enough bandwidth for the assigned flows:
\begin{equation}
   \label{eq:bandwidth}
   \begin{split}
        &\forall p \in P,\forall q \in p.Q, \forall f \in \mathcal{F}(q): \frac{q.w}{q.T}\geq \sum \frac{f.l}{f.T}.
   \end{split}
\end{equation}
where $\mathcal{F}(q)$ is the set of flows assigned to the queue~$q$.

\noindent (4) The \textbf{Port Period Constraint} (Eq.~\eqref{eq:sameportperiods}) imposes that the periods of all the queues in a port should be harmonic. This constraint is used to avoid window overlapping as well as reducing the search space.
\begin{equation}
   \label{eq:sameportperiods}
   \begin{split}
        &\forall p \in P,\forall q \in p.Q, \forall q' \in p.Q: (q.T\%q'.T=0) \vee (q'.T\%q.T=0).
   \end{split}
\end{equation}
(5) The \textbf{Period Limit Constraint} (Eq.~\eqref{eq:fixedperiods}) reduces the search space by considering window periods~$q.T$ that are harmonic with the hyperperiod of the port $\mathcal{K}(p)$ (divide it):
\begin{equation}
   \label{eq:fixedperiods}
   \begin{split}
        &\forall p \in P,\forall q \in p.Q: \quad \mathcal{K}(p)\%{q.T}= 0.
   \end{split}
\end{equation}

\subsection{Timing constraints}
\label{sec:proxy}

\begin{figure}
     \centering
     \begin{subfigure}[b]{0.45\textwidth}
         \centering
         \includegraphics[width=\textwidth]{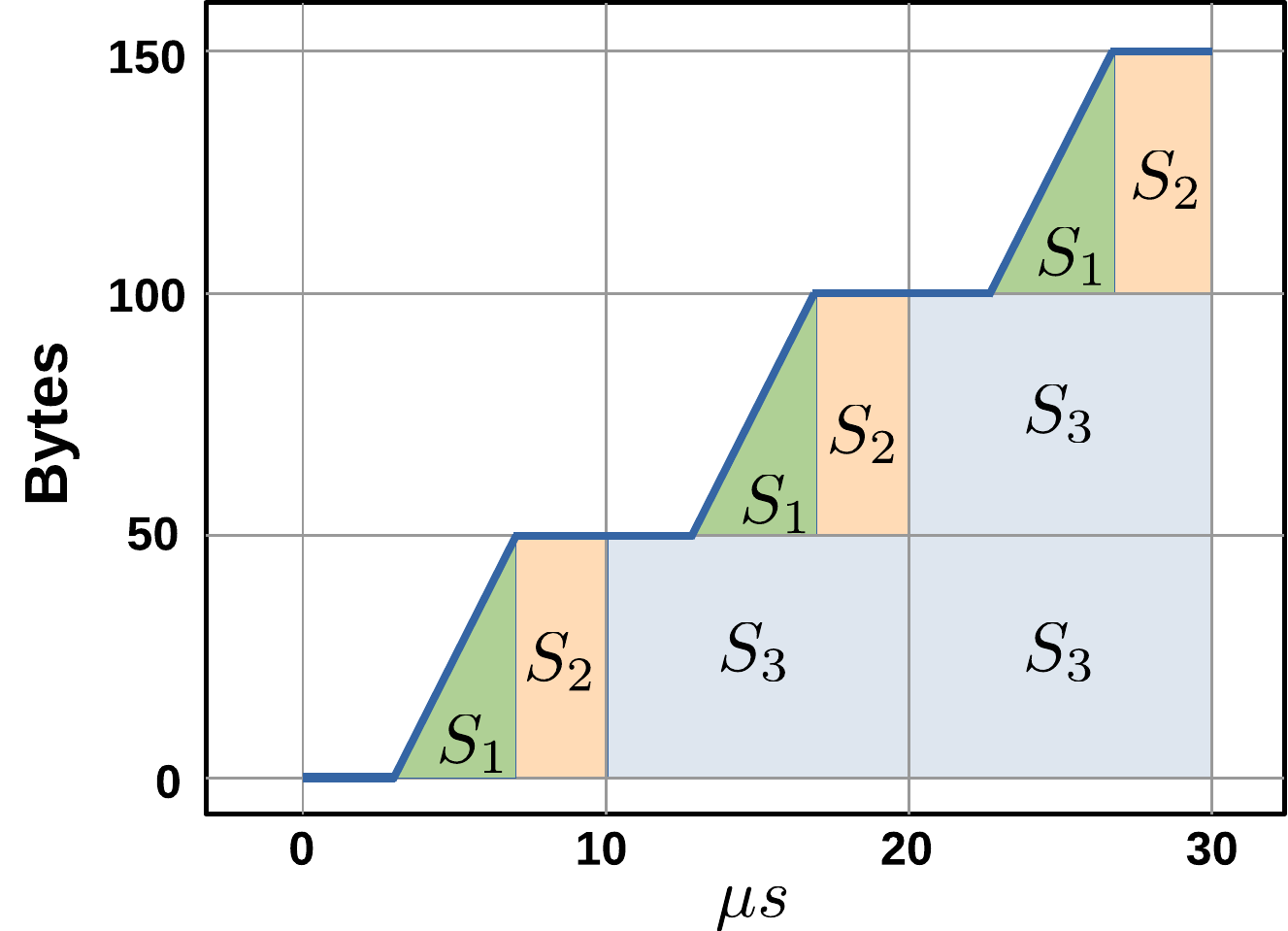}
         \caption{example capacity}
         \label{fig:service}
     \end{subfigure}
     \hfill
     \begin{subfigure}[b]{0.47\textwidth}
         \centering
         \includegraphics[width=\textwidth]{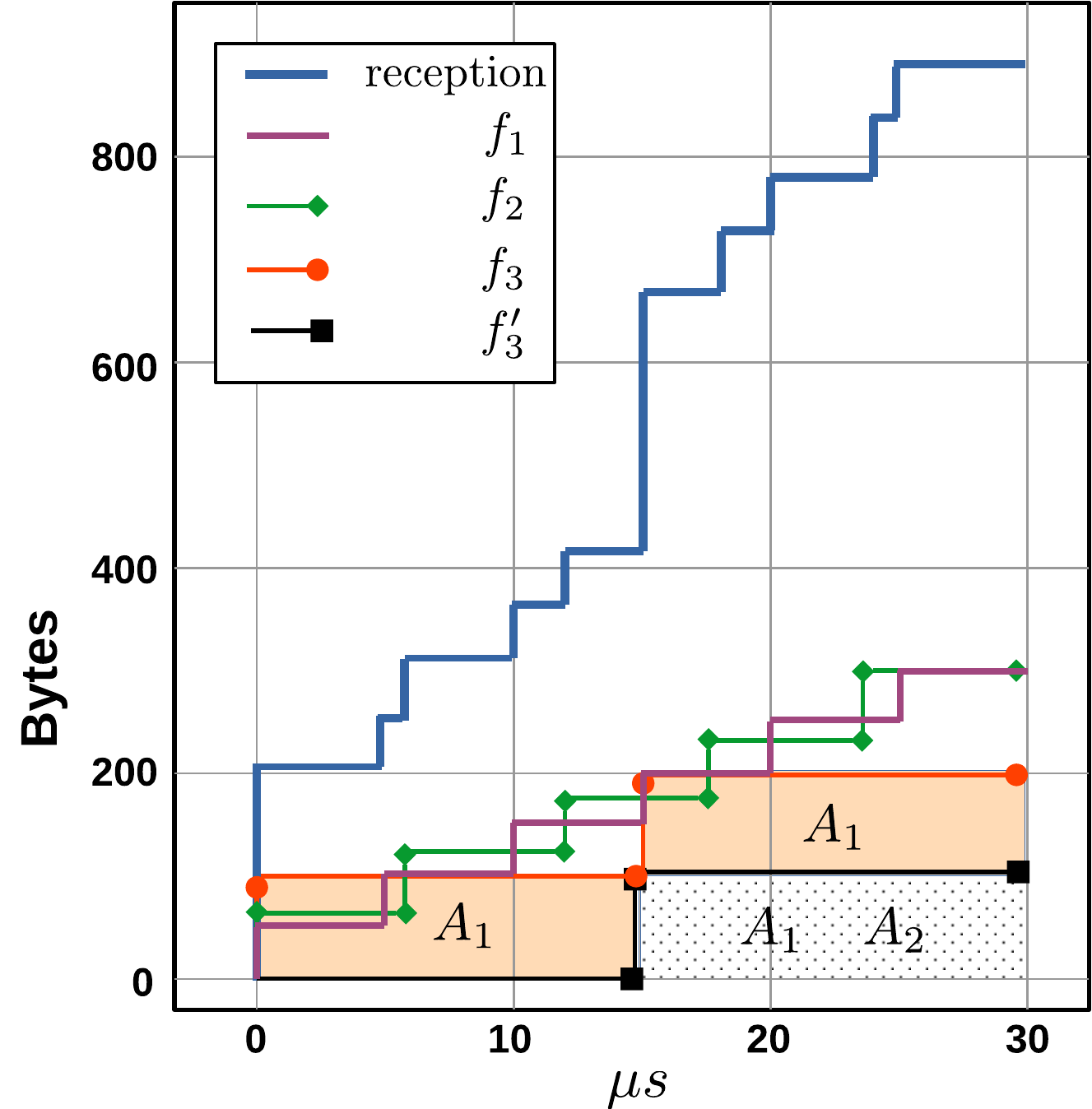}
         \caption{example transmission demand}
         \label{fig:reception}
     \end{subfigure}
     \caption{Example capacity and transmission demand for a window.}
\end{figure}

As mentioned, it is infeasible to use a Network Calculus-based worst-case delay analysis to check the schedulability of \textit{each} solution visited. Thus, we have defined a \emph{Timing Constraint} as a way to prune the search space. Every solution that is not eliminated via this timing constraint is evaluated for schedulability with the NC WCD analysis. The timing constraint is a heuristic that prunes the search space of (potentially unschedulable) solutions; it is not a sufficient nor a necessary schedulability test. The timing constraint is related to the optimality of the solution, not to its correctness in terms of schedulability. A too aggressive pruning may eliminate good quality solutions, and too little pruning will slow down the search because the NC WCD analysis is invoked too often.

The challenge is that the min+ algebra used by NC cannot be directly expressed in first-order formulation of CP. However, the NC formulation from~\cite{Zhao20:JIoT} has inspired us in defining the CP timing constraints. The \emph{Timing Constraint} is defined in Eq.~\eqref{eq:deadlineConstraint} and uses the concepts of \textit{window capacity} $\mathcal{W}_C$ and \textit{transmission demand} $\mathcal{W}_D$ to direct the CP solver to visit only those solutions where the \textit{capacity} of each window, i.e., the amount of time available to transmit frames assigned to its queue, is greater than or equal to its transmission \textit{demand}, i.e., the amount of transmission time required by the frames in the queue. A window capacity larger than the transmission demand indicates that a solution has high chances to be schedulable:
\begin{equation}
   \label{eq:deadlineConstraint}
   \begin{split}
        &\forall p \in P,\forall q \in p.Q: \mathcal{W}_D \leq \mathcal{W}_C.
   \end{split}
\end{equation}

\noindent Thus, we first calculate the capacity $\mathcal{W}_C$ of each window within the hyperperiod. This capacity is similar to the NC concept of a service curve, and its calculation is similar to the service curves proposed in the literature~\cite{wandeler2006modular} for resources that use Time-Division Multiple Access (TDMA), which is how our windows behave. For e.g., a window with a period of 10~$\mu$s, a length of 4~$\mu$s, and an offset of 3~$\mu$s; forwards 150 bytes over a 100 Mbps link in a hyperperiod of 30~$\mu$s. In Fig.~\ref{fig:service}, the capacity of such a window is depicted where the blue line shows the throughput of the window for transferring data. The capacity increases when the window opens (the rising slopes of the curve). The effect of window offset on the capacity (the area under the curve) can be observed in the figure.
The function~$\mathcal{W}_C$ calculates the area under the curve to characterize the amount of capacity for a window in a hyperperiod, defined in Eq.~\eqref{eq:serviceVal}, where the link~$[v_a, v_b]$ is attached to the port~$p$ and assigned to the queue~$q$; and function $\mathcal{Y}$ captures the transmission time of a single byte through link~$[v_a, v_b]$.

To calculate the area under the curve, we consider $3$ terms that are $S_1$, $S_2$, and $S_3$. They represent the total area under the curve caused by the window length, the window closure in the remainder of the window period, and the window period, respectively. 
The $\mathcal{W}_C$ value of the example in Fig.~\ref{fig:service} is 2,250 $Bytes\times \mu s$, where the $S$ terms are shown.
\begin{equation}
   \label{eq:serviceVal}
   \begin{split}
        &\forall p \in P,\forall q \in p.Q:\\
        &I = \frac{\mathcal{K}(p)}{q.T}, \quad J=\frac{(q.w - \mathcal{GB}(q))\times\mathcal{Y}([v_a, v_b])}{[v_a, v_b].C},\\
        &S_1 = I \times \frac{q.w \times J}{2}, \quad S_2= I \times (q.T - q.w. - q.\phi) \times J,\\
        &S_3=\frac{I\times (I-1)}{2} \times q.T \times J, \quad \mathcal{W}_C=S_1 + S_2+S_3
   \end{split}
\end{equation}

Secondly, we calculate the transmission demand $\mathcal{W}_D$ using Eq.~\eqref{eq:receptionVal}, where the function~$\mathcal{R}(q)$ captures all the flows that are assigned to the queue~$q$. The transmission demand is inspired by the \textit{arrival curves} of NC. These are carefully determined in NC considering that the flows pass via switches and may change their arrival patterns~\cite{Zhao20:JIoT}. In our case, we have made the following simplifying assumptions in order to be able to express the ``transmission demand'' in CP. We assume that all flows are strictly periodic and arrive at the beginning of their respective periods. This is ``optimistic'' with respect to NC in the sense that NC may determine that some of the flows have a bursty behavior when they reach our window. To compensate for this, we consider that those flows that arrive from a switch may be bursty and thus have a backlog $\mathcal{B}$ of frames that have accumulated; flows that arrive from ESs do not accumulate a backlog. Fig.~\ref{fig:reception} shows three flows, $f_1$ to $f_3$, and only $f_3$ arrives from a switch and hence will have a backlog of frames captured by the flow denoted with $f'_3$ (we consider a $\mathcal{B}$ of 1 in the example). We also assume that the backlog $f'_3$ will not arrive at the same time as the original flow $f_3$, and instead, it is delayed by a period. Again, this is a heuristic used for pruning, and the actual schedulability check is done with the NC analysis. So, the definition of the ``transmission demand'' does not impact correctness, but, as discussed, it will impact our algorithm's ability to search for solutions. 

Since, in our case, the deadlines can be larger than the periods, we also need to consider for each flow bursts of frames coming from SWs and an additional frame for each flow coming from ESs (the ES periods are not synchronized with the SWs GCLs). Since we do not perform a worst-case analysis, we instead use a backlog parameter $\mathcal{B}$, which captures the possible number of delayed frames in a burst within a flow forwarded from another SW. Note that as explained in the overview at the beginning of Sect.~\ref{sec:solution}, $\mathcal{B}$ is a user-defined parameter that controls the ``pruning level'' of our timing constraint, i.e., how aggressively it eliminates candidates from the search space.

We give an example in Fig.~\ref{fig:reception} where the flows~$f_1<50,5,0,5>$ and $f_2<60,6,0,6>$ have been received from an ES and the flow~$f_3<100,15,0,15>$ has been received from a SW. For the flow~$f_3$ forwarded from a previous switch, we consider that one instance of the flow (determined by the backlog parameter $\mathcal{B} = 1$), let us call it~$f'_3$, may have been delayed and received together with the current instance~$f_3$. This would cause a delay in the reception of the flows in the current node. The reception curve in Fig.~\ref{fig:reception} is the sum of curves for each stream separately in a hyperperiod of 30~$\mu$s. 

We give the general definition of the transmission demand value $\mathcal{W}_D$ as the area under the curve for the accumulated data amount of received flows and backlogs of the flows arrived from switches in a hyperperiod. For calculating the transmission demand $\mathcal{W}_D$, we consider 2 terms that are $A_1$ and $A_2$.
The term $A_1$ calculates the area under the curve for the accumulated data of all flows assigned to the queue~$q$ captured by~$\mathcal{R}(q)$, in a hyperperiod. Any frames of all flows~$\mathcal{R}(q)$ have arrived at the beginning of their period. The term~$A_2$ calculates the area under the curve for the accumulated backlog data of the flows arrived from a switch captured by~$\mathcal{X}(q)$. The backlog data of the flows~$\mathcal{X}(q)$ are delayed for a period and controlled by~$\mathcal{B}$, which captures the number of backlogs.
The function $\mathcal{W}_D$ returns 16,650 $Bytes\times \mu s$ in our example, see also Fig.~\ref{fig:reception} for the values of the terms $A_1$ and  $A_2$.
\begin{equation}
   \label{eq:receptionVal}
   \begin{split}
        &\forall p \in P,\forall q \in p.Q, \forall f \in \mathcal{R}(q), \forall f' \in \mathcal{X}(q):\\
        &I = \frac{\mathcal{K}(p)}{f.T}, \quad I' = \frac{\mathcal{K}(p)}{f'.T}\\
        &A_1=\frac{I \times (I+1)}{2} \times f.T \times f.l,\\
        &A_2=\frac{I' \times (I'+1 - 2\times \mathcal{B})}{2} \times f'.T \times f'.l,\\
        &\mathcal{W}_D= A_1 + A_2
   \end{split}
\end{equation}

Please note that the correctness of the constraints (Eq.~(\ref{eq:windowproperty}), (\ref{eq:overlap}), (\ref{eq:bandwidth})) follows from the implicit hardware constraints of 802.1Q(bv) (see the discussion in~\cite{Craciunas16:RTNS, Oliver18}) while other constraints (Eq.~(\ref{eq:sameportperiods}), (\ref{eq:fixedperiods})) are used to limit the placement of GCL windows and are not related to correctness, just to optimality. Since the transmission of frames is decoupled from the GCL windows, the schedule's correctness concerning the end-to-end latency of streams is always guaranteed due to the NC analysis, which is intertwined in the schedule step.

\section{Evaluation}
\label{sec:Evaluation}
In this section, we first discuss the ST scheduling approaches in the literature, motivating our proposed approach called Flexible Window-based GCL (\emph{FWND}), see Sect.~\ref{sec:approaches}. We give details of our setup and test cases in Sect.~\ref{sec:setup} and evaluate our windows optimization solution CPWO for \emph{FWND} on synthetic and real-world test cases on Sect.~\ref{sec:syntestcases} and Sect.~\ref{sec:realisticcases}, respectively. We also compare the CPWO results with the related work and validate the generated GCLs with OMNET++ in Sect.~\ref{sec:omnet}.

\subsection{Scheduling Approaches}
\label{sec:approaches}
The related work on ST scheduling using 802.1Qbv consists of: (i)~zero-jitter GCL (0GCL)~\cite{Craciunas16:RTNS, Pop16}, (ii)~Frame-to-Window-based GCL (FGCL)~\cite{Oliver18}, and (iii)~Window-based GCL (WND)~\cite{ReuschWFCS20}. 

\begin{table}[!b]
\newcommand{\tabincell}[2]{\begin{tabular}{@{}#1@{}}#2\end{tabular}}
\small
\centering
\scalebox{0.99}{
\begin{threeparttable}[b]
\caption{Time-Sensitive Message Transmission Approaches in TSN}
\label{tab:problemcompositions}
\centering{
\begin{tabular}{l|c|c|c|c}
\hline
\textbf{Requirements} &\tabincell{c}{\textbf{0GCL}\\\cite{Craciunas16:RTNS}\cite{Pop16}}  & \tabincell{c}{\textbf{FGCL}\\\cite{Oliver18}}  & \tabincell{c}{\textbf{WND}\\\cite{ReuschWFCS20}}& \tabincell{c}{\textbf{FWND}}\\
\hline
\hline
Device Capabilities  & \tabincell{c}{802.1Qbv}  & \tabincell{c}{802.1Qbv}  & \tabincell{c}{802.1Qbv} & \tabincell{c}{802.1Qbv}\\
ES Capabilities  & scheduled  & scheduled  & non-scheduled & non-scheduled\\
SW Capabilities & scheduled  & scheduled  & scheduled & scheduled \\
\hline
Frame Constraint  & Yes  & Yes  & Yes & Yes\\
Link Constraint\tnote{1}  & Yes  & Yes  & No & No \\
Bandwidth Constraint & No & No & No & Yes   \\
\tabincell{c}{Flow Transmission Constraint\tnote{2}}  & Yes  & Yes  & No & No\\
\tabincell{c}{Frame-to-Window Assignment}  & No  & Yes  & No & No \\
Window Size Constraint  & No  & Yes  & Yes & Yes\\
Flow/Frame Isolation  & Yes  & Yes  & No & No\\
End-to-end Constraint  & Yes  & Yes  & Yes & Yes \\
\hline
Schedule synthesis  &  \tabincell{c}{Yes \\(intractable)} &  \tabincell{c}{Yes \\(intractable)} & \tabincell{c}{No (only\\windows)} & \tabincell{c}{No (only\\windows)}\\
Timing analysis required  & No  & No  & Yes & Yes\\
\hline
\end{tabular}}
\begin{tablenotes}
\item[1] No two frames routed through the same physical link can overlap in the time domain, also named ``Ordered Windows Constraint'' in~\cite{Oliver18}.
\item[2] The propagation of frames of a flow follows the sequential order along the path of the flow, also named ``Stream Constraint'' in~\cite{Oliver18}.
\end{tablenotes}
\end{threeparttable}}
\end{table}

We summarize the requirements of the ST scheduling approaches from the related work and our \emph{FWND} approach in the first column of Table~\ref{tab:problemcompositions}.
The first three requirements refer to the device capabilities needed for the different approaches, and the next seven rows present the constraints and isolation requirements.
The last two rows present the requirements of the complexity of the optimization problem that needs to be solved to provide a solution for the respective approach. 

To better understand the fundamental differences and the similarities (in terms of the imposed correctness constraints and schedulability parameters) between our work and the approaches that require synchronized and scheduled end systems, we briefly reiterate the formal constraints of previous work. We describe, based on~\cite{Craciunas16:RTNS, CraciunasETR17, Oliver18}, the relevant scheduling constraints for creating correct TSN schedules when using frame- and window-based methods. Table~\ref{tab:problemcompositions} shows which of these are needed by which approach. We adapt some notations from~\cite{Craciunas16:RTNS, CraciunasETR17, CraciunasJRTS16} to describe the constraints and assume certain simplifications without loss of generality, e.g. the macrotick is the same in all devices, all flows have only one frame per period, the propagation delay $d_p$ is $0$. We refer the reader to~\cite{Craciunas16:RTNS, Oliver18} for a complete and generalized formal definition of the correctness constraints. We denote the messages (frames) of a flow $f_i$ on a link $[v_a, v_b]$ as $m_i^{[v_a, v_b]}$. A message $m_i^{[v_a, v_b]}$ is defined by the tuple $\langle m_i^{[v_a, v_b]}.\phi, m_i^{[v_a, v_b]}.l \rangle,$ denoting the transmission time and duration of the frame on the respective link~\cite{Craciunas16:RTNS, CraciunasETR17}. 

\textbf{Frame Constraint.}
Any frame belonging to a critical flow has to be transmitted between time $0$ and its period $T_i$. To enforce this, we have the frame constraint from~\cite{Craciunas16:RTNS}:
\begin{align}
&\forall f_i \in \mathcal{F}, \forall [v_a, v_b] \in f_i.r :
\left(m_i^{[v_a, v_b]}.\phi \ge 0\right) \wedge \left( m_i^{[v_a, v_b]}.\phi \le f_i.T - m_i^{[v_a, v_b]}.l \right). \nonumber
\end{align} 

\textbf{Link Constraint.}
Since there can only be one frame at a time on a physical link, we enforce that no two frames sent on the same egress port may overlap in the time domain. The link constraint adapted from~\cite{Craciunas16:RTNS} is hence: 
\begin{align}
&\forall [v_a, v_b] \in \bm{E}, \forall m_i^{[v_a, v_b]}, m_j^{[v_a, v_b]} (i \neq j), \forall a \in [0, hp_{i}^{j}/f_i.T - 1 ], \forall b \in [0, hp_i^j/f_j.T - 1]:\nonumber\\
&\Bigl(m_i^{[v_a, v_b]}.\phi + a\times f_i.T \ge
m_j^{[v_a, v_b]}.\phi + b\times f_j.T + m_j^{[v_a, v_b]}.l \Bigr)\vee  \nonumber \\&  \Bigl(m_j^{[v_a, v_b]}.\phi + b\times f_j.T \ge 
m_i^{[v_a, v_b]}.\phi + a\times f_i.T + m_i^{[v_a, v_b]}.l\Bigr),\nonumber
\end{align}
where $hp_i^j = lcm(f_i.T, f_j.T)$ is the hyperperiod of $f_i$ and $f_j$.

\textbf{Flow Transmission Constraint.}
The (optional) flow transmission constraint enforces that a frame is forwarded by a device only after it has been received at that device also taking into account the network precision, denoted with $\delta$: 
$$\forall f_i \in \mathcal{F}, \forall [v_a, v_x], [v_x, v_b] \in f_i.r, \forall m_i^{[v_a, v_x]}, \forall m_i^{[v_x, v_b]}: m_i^{[v_x, v_b]}.\phi  - \delta \ge 
m_i^{[v_a, v_x]}.\phi + m_i^{[v_a, v_x]}.l.$$

\textbf{End-to-End Constraint.}
The maximum end-to-end latency constraint (expressed by the deadline $f_i.D$) enforces a maximum time between the sending and the reception of a flow. We denote the sending 
link of stream $f_i$ with $src(f_i)$ and the last link before the receiving node with $dest(f_i)$. 
The maximum end-to-end latency constraint~\cite{Craciunas16:RTNS} is hence 
\begin{align}
&\forall f_i \in \mathcal{F}: m_i^{dest(f_i)}.\phi + m_i^{dest(f_i)}.L - m_i^{src(f_i)}.\phi \le f_i.D - \delta \nonumber.
\end{align}
Here again the network precision $\delta$ needs to be taken into account since the local times of the sending and receiving devices can deviate by at most $\delta$.

\textbf{802.1Qbv Flow/Frame Isolation.} 
Due to the non-determinism problem in TSN (c.f. Sect.~\ref{sec:sw-model}), previous solutions (e.g.,~\cite{Craciunas16:RTNS, Oliver18}) need an isolation constraint that maintains queue determinism. We refer the reader to~\cite{Craciunas16:RTNS} for an in-depth explanation and only summarize here the flow and frame isolation constraint adapted from~\cite{Craciunas16:RTNS}. Let $m_i^{[v_a, v_b]}$ and $m_j^{[v_a, v_b]}$ be, respectively, the frame instances of $f_i \in \mathcal{F}$ and $f_j \in \mathcal{F}$ scheduled on the outgoing link $[v_a, v_b]$ of device $v_a$. Flow $f_i$ arrives at the device $v_a$ from some device $v_x$ on link $[v_x, v_a]$. Similarly, flow $f_j$ arrives from another device $v_y$ on incoming link $[v_y, v_a]$. The simplified flow isolation constraint adapted from~\cite{Craciunas16:RTNS}, under the assumption that the macrotick of the involved devices is the same, is as follows:
\begin{align}
&\forall [v_a, v_b] \in \bm{E}, \forall m_i^{[v_a, v_b]}, m_j^{[v_a, v_b]} (i\neq j), \forall a \in [0, hp_{i}^{j}/f_i.T - 1 ], \forall b \in [0, hp_i^j/f_j.T - 1]:\nonumber\\
&\Bigl(m_i^{[v_a, v_b]}.\phi + a\times f_i.T + \delta \le 
m_j^{[v_y, v_a]}.\phi + b\times f_j.T \Bigr) \vee \nonumber\\
&\Bigl(m_j^{[v_a, v_b]}.\phi + b\times f_j.T + \delta \le 
m_i^{[v_x, v_a]}.\phi + a\times f_i.T \Bigr). \nonumber
\end{align}
Here again $hp_i^j = lcm(f_i.T, f_j.T)$ is the hyperperiod of $f_i$ and $f_j$. The constraint ensures that once a flow arrives at a device, no other flow can enter the device until the first flow has been sent. 

The above constraints apply to frames that are placed in the same queue on the egress port. However, the scheduler may choose (if possible) to place streams in different queues, isolating them in the space domain.
Hence, the complete constraint~\cite{Craciunas16:RTNS} for frame/flow isolation for two flows $f_i$ and $f_j$ scheduled on the same link $[v_a, v_b]$ can be expressed as  
\begin{align}
&\Bigl( \Phi_{[v_a, v_b]}(f_i, f_j) \Bigr) \vee \Bigl(m_i^{[v_a, v_b]}.q \neq m_j^{[v_a, v_b]}.q \Bigr), \nonumber
\end{align}
with $m_i^{[v_a, v_b]}.q \le N_{tt}$ and $m_j^{[v_a, v_b]}.q \le N_{tt}$ and where $\Phi_{[v_a, v_b]}(f_i, f_j)$ denotes either the flow or frame isolation constraint from before.

\textbf{Decoupling of frames.}
So far, the constraints were applicable on the level of frames, and the open windows of the GCLs were constructed from the resulting frame schedule. The approach in~\cite{Oliver18} decouples the frame transmission from the respective open gate windows defined in the GCLs, similar to our approach\footnote{Note that ~\cite{Craciunas16:RTNS, Oliver18, Pop16} cannot be used in our context because they require scheduled and synchronized ESs}. However, in~\cite{Oliver18} the requirement is that there is a unique assignment of which frames are transmitted in which windows, although also multiple frames can be assigned to be sent in the same window. Hence, the assignment of frames and, consequently, the length of each gate open window are, therefore, an output of the scheduler.  Therefore, we have to construct additional constraints when (partially) decoupling frames from windows. For a more in-depth description and formalization of these constraints we refer the reader to~\cite{Oliver18}.

\textit{Frame-to-Window Assignment.} The frame-to-window assignment restricts a frame to be assigned to a specific window, although multiple frames can be assigned to the same window.

\textit{Window Size Constraint.} The length of the gate open window has to be restricted to the sum of the frame lengths that have been assigned to it. 

Comparing the existing approaches with the one proposed in this paper, we see that the choice of scheduling mechanism is, on the one hand, highly use-case specific and, on the other hand, is constrained by the available TSN hardware capabilities in the network nodes. While the frame- and window-based methods from related work result in precise schedules that emulate either a 0- or constrained-jitter approach (e.g., like in TTEthernet), they require end systems to not only be synchronized to the network time but also the end devices to have 802.1Qbv capabilities, i.e., to be scheduled. This limitation might be too restrictive for many real-world systems relying on off-the-shelf sensors, processing, and actuating nodes. While our \emph{FWND} method overcomes this limitation, it does require a worst-case end-to-end analysis that introduces a level of pessimism into the timing bounds, thereby reducing the schedulability space for some use cases. However, as seen in Table~\ref{tab:problemcompositions}, our method does not require many of the constraints imposed on the flows and scheduled devices from previous work, thereby reducing the complexity of the schedule synthesis.

\subsection{Test Cases and Setup}
\label{sec:setup}
\emph{FWND} is implemented in Java using Google OR-Tools~\cite{ortools} and the Java kernel of the RTC toolbox~\cite{Wanderler06} and was run on a computer with an i9 CPU at 3.6 GHz and 32 GB of RAM. We have considered a time limit of 10 to 90 minutes, depending on the test case size. The macrotick and $\mathcal{B}$ parameters are set to 1 $\mu$s and 1, respectively, in all the test cases.

\begin{figure}[t]{}
	\centering
	\includegraphics[width=0.99\textwidth]{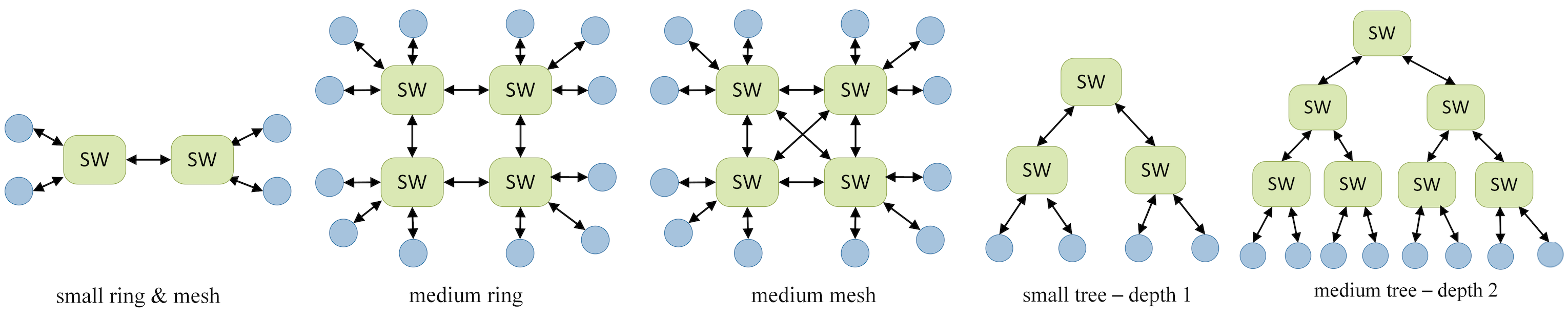}
	\caption{Network topologies used in the test cases}
	\label{fig:SynTC_topo}
\end{figure}

We have generated 15 synthetic test cases that have different network topologies (three test cases for each topology in Fig.~\ref{fig:SynTC_topo}) inspired by industrial and automotive application requirements. Similar to~\cite{zhao2021quantitative}, the network topologies are small ring \& mesh (SRM), medium ring (MR), medium mesh (MM), small tree-depth 1 (ST), and medium tree-depth 2 (MT).
Flows of the test cases are randomly generated, and the message sizes are randomly chosen between the minimum and the maximum Ethernet frame size, while their periods are selected from the set $P=\{\textrm{1,500, 2,500, 3,500, 5,000, 7,500, 10,000}\}$, all in $\mu$s. The physical link speed is set for 100 Mbps. The details of the synthetic test cases are in Table~\ref{tab:synthticcases} where the second column shows the topology of the test cases, and the number of switches, end systems, and flows are shown in columns 3 to 6.

We have also used two realistic test cases: an automotive case from General Motors (GM) and an aerospace case, the Orion Crew Exploration Vehicle (CEV). The GM case consists of 27 flows varying in size between 100 and 1,500 bytes, with periods between 1~$ms$ and 40~$ms$ and deadlines smaller or equal to the respective periods. The CEV case is larger, consisting of 137 flows, with sizes ranging from 87 to 1,527 bytes, periods between 4~$ms$ and 375~$ms$, and deadlines smaller or equal to the respective periods. The physical link speed is set for 1000 Mbps. More information can be found in the corresponding columns in Table~\ref{tab:test_cases_large}. Use cases use the same topologies as in~\cite{Gavrilut17} and~\cite{Zhao20:JIoT}, and we consider that all flows are ST. 

\begin{table}[t]
\small
\caption{Details of the synthetic test cases}
\label{tab:synthticcases}
\centering
\scalebox{0.79}{
\begin{adjustbox}{width=0.99\textwidth}
\small
\begin{tabular}{l|c|c|c|c|c}
\hline
\textbf{No.}&\textbf{Network} &\textbf{Total No.} &\textbf{Total No.} &\textbf{Total No.} &\textbf{Hyperperiod}\\
\textbf{}  &\textbf{Topology}&\textbf{of SWs}&\textbf{of ESs}&\textbf{of Flows}& \textbf{($\mu$s)}\\
\hline
\hline
1  & SRM  & 2   & 3     & 9     & 15,000\\
2  & SRM  & 3   & 3     & 11    & 70,000\\
3  & SRM  & 3   & 4     & 15    & 70,000\\
4  & MR  & 4   & 6     & 15     & 30,000\\
5  & MR  & 4   & 8     & 21     & 210,000\\
6  & MR  & 5   & 11     & 27    &210,000\\
7  & MM  & 4   & 5     & 13     & 15,000\\
8  & MM  & 6  & 12     & 30& 210,000\\
9  & MM  & 7   & 13     & 35& 210,000\\
10  & ST  & 3   & 4     & 7& 15,000\\
11  & ST  & 3   & 6     & 12& 15,000\\
12  & ST  & 3   & 7     & 16& 105,000\\
13  & MT  & 7   & 8     & 18& 105,000\\
14  & MT  & 7   & 8     & 25& 105,000\\
15  & MT  & 7  & 12     & 32& 210,000\\
\hline
\end{tabular}
\end{adjustbox}
}
\end{table}

\subsection{Evaluation on synthetic test cases}
\label{sec:syntestcases}

We have evaluated our CPWO solution for \emph{FWND} on synthetic test cases. The results are depicted in Table~\ref{tab:evalsynthetic} where we show the objective function value (average bandwidth $\Omega$ from Eq.~(\ref{eq:avgbandwidth})) and the mean WCDs. For quantitative comparison, we have also reported the results for the three other ST scheduling approaches: \emph{0GCL}, \emph{FGCL}, \emph{WND}. \emph{0GCL} and \emph{FGCL} were implemented by us with a CP formulation using the constraints from~\cite{Craciunas16:RTNS} and~\cite{Oliver18}, respectively, where these techniques were proposed. The \emph{WND} method has been implemented with the heuristic presented in~\cite{ReuschWFCS20}, but instead of using the WCD analysis from~\cite{Zhao18:Access}, we extend it to use the analysis from~\cite{Zhao20:JIoT} instead, in order not to unfairly disadvantage WND over our CPWO solution. Note that the respective mean worst-case end-to-end delays in the table are obtained over all the flows in a test case, from a single run of the algorithms, since the output of the algorithms is deterministic based on worst-case analyses, not based on simulations.

It is important to note that \emph{0GCL} and \emph{FGCL} are presented here as a means to evaluate CPWO; however, they are \textit{not producing valid solutions} for our problem, which considers unscheduled end systems, see Table~\ref{tab:problemcompositions} for the requirements of each method. As expected, when end systems are scheduled and synchronized with the rest of the network as is considered in \emph{0GCL} and \emph{FGCL}, we obtain the best results in terms of bandwidth usage ($\Omega$) and WCDs, with the comment that \emph{0GCL} may further reduce the WCDs compared to \emph{FGCL}.

The only other approach that has similar assumptions to our \emph{FWND} is \emph{WND} from~\cite{ReuschWFCS20}. As we can see from Table~\ref{tab:evalsynthetic}, in comparison to \emph{WND}, our CPWO solution can slightly reduce the bandwidth usage. The most important result is that CPWO significantly reduces the WCDs compared to \emph{WND}, with an average of 104\% and up to 437\% for some test cases such as TC13. This means that we are able to obtain schedulable solutions in more cases compared to the work in~\cite{ReuschWFCS20}. 

Also, when comparing the WCDs obtained by our \emph{FWND} approach with the case when the end systems are scheduled, i.e., \emph{0GCL} and \emph{FGCL}, we can see that the increase in WCDs is not dramatic. This means that for many classes of applications, which can tolerate a slight increase in latency, we can use our CPWO approach to provide solutions for more types of network implementations, including those that have unscheduled and/or unsynchronized end systems. In addition, due to the complexity of their CP model, it takes a long runtime to obtain solutions for \emph{0GCL} and \emph{FGCL}, and the CP-model for \emph{FGCL} run out of memory for some of the test cases (the NA in the table). As shown in the last two columns of Table~\ref{tab:evalsynthetic}, where we present the runtimes of \emph{0GCL} and \emph{FWND}, \emph{FWND} reduces the runtime significantly. The reason for reduced runtime with \emph{FWND} is that the CP model has to determine values for fewer variables compared to \emph{0GCL}. \emph{FWND} introduces 3 variables (offset, period and length) for each window (queue) in the network, whereas \emph{0GCL} introduces a variable for each frame of each flow. The number of variables in the \emph{0GCL} model depends on the hyperperiod, the number of flows, and the flow periods, whereas the number of variables in the \emph{FWND} model depends on the number of switches and used queues. 

\begin{table}[b]
\caption{Performance of CPWO for FWND on realistic test cases}
\label{tab:test_cases_large}
\centering{
\begin{tabular}{l|c|c}
\hline                               
                        & \textbf{ORION (CEV)}&\textbf{GM}\\
\hline
ES                      & 31            & 20\\
SW                      & 15            & 20\\
Flows                   & 137           & 27\\
\hline
Mean WCDs ($\mu$s) & 10,376           & 1,981\\
$\Omega$ ($\times$1000) & 435          & 84 \\
Runtime (s)        & 891   & 17\\
\hline
\end{tabular}}
\end{table}

\begin{table*}[t!]
\caption{Evaluation results on synthetic test cases}
\label{tab:evalsynthetic}
\centering
\begin{adjustbox}{width=0.99\textwidth}
\small
\begin{threeparttable}[b]
\begin{tabular}{l|c|c|c|c|c|c|c|c|c|c}
\hline
\textbf{No.}&\textbf{$\Omega$\tnote{1}}&\textbf{$\Omega$\tnote{1}}&\textbf{$\Omega$\tnote{1}}&\textbf{$\Omega$\tnote{1}}&\textbf{Mean worst-case}&\textbf{Mean worst-case}&\textbf{Mean worst-case}&\textbf{Mean worst-case}&\textbf{Mean Runtime}&\textbf{Mean Runtime}\\
\textbf{}  &\textbf{for}     &\textbf{for}     &\textbf{for}     &\textbf{for}     &\textbf{e2e-delay for}&\textbf{e2e-delay for}&\textbf{e2e-delay for}&\textbf{e2e-delay for}&\textbf{for}&\textbf{for}\\
           &\textbf{0GCL}   &\textbf{FGCL} &\textbf{WND} &\textbf{FWND} &\textbf{0GCL ($\mu$s)}     &\textbf{FGCL($\mu$s)}  &\textbf{WND ($\mu$s)}  &\textbf{FWND ($\mu$s)}&\textbf{0GCL (ms)} &\textbf{FWND (ms)}\\
\hline
1   &35     &35    &614     &510   &192     &126     &1838     &1556 &215    &8\\
2   &25     &22    &640     &528   &246     &151     &2461     &1806    &895    &12\\
3   &15     &15   &549     &495   &175     &486     &1964     &1384    &1518   &22\\
4   &13     &13    &330     &285   &160     &776     &2925     &1832 &525    &16\\
5   &14     &NA\tnote{2}     &295     &285   &131     &NA\tnote{2}     &2838     &2347 &5187   &16\\
6   &13     &NA\tnote{2}     &275     &205   &129     &NA\tnote{2}     &2953     &1976  &6291   &17\\
7   &12     &12    &238     &204   &125     &764     &2913     &1561 &1152   &35\\
8   &13     &NA\tnote{2}     &238     &202   &114     &NA\tnote{2}     &2878     &1725  &7603   &36\\
9   &12     &NA\tnote{2}     &217     &191   &122     &NA\tnote{2}     &3074     &1927  &9171   &56\\
10  &8      &8     &329     &265   &136     &2284     &4397     &4327    &2611   &165\\
11  &10      &10     &381     &302   &159     &984     &3047     &2057 &2840   &231\\
12  &11     &NA\tnote{2}     &516     &321   &187     &NA\tnote{2}     &2543     &1326    &4650   &260\\
13  &10      &10     &401     &302   &101     &561     &2529     &471 &978    &130\\
14  &9      &9     &611     &402   &120     &785     &2254     &628 &1256   &162\\
15  &9      &NA\tnote{2}     &544     &413   &114     &NA\tnote{2}      &2680     &713 &6116   &163\\
\hline
\end{tabular}
\begin{tablenotes}
\item[1] Values are multiplied by 1000
\item[2] Ran out of memory
\end{tablenotes}
\end{threeparttable}
\end{adjustbox}
\end{table*}

\subsection{Evaluation on realistic test cases}
\label{sec:realisticcases}
We have used two realistic test cases to investigate the scalability of CPWO and its ability to produce schedulable solutions for real-life applications. The results of the evaluation are presented in Table~\ref{tab:test_cases_large} where the mean WCDs, objective value $\Omega$, and runtime for the two test cases are given. As we can see, CPWO has successfully scheduled all the flows in both test cases. Note that once all flows are schedulable, CPWO aims at minimizing the bandwidth. This means that CPWO may be able to achieve even smaller WCD values at the expense of bandwidth usage. In terms o runtime, the CEV test case takes longer since it has $864$ variables, whereas GM has only $102$ variables in the CP models.

\subsection{Validating the solutions with OMNET++}
\label{sec:omnet}
\begin{figure}
     \centering
     \begin{subfigure}[b]{0.45\textwidth}
         \centering
         \includegraphics[width=\textwidth]{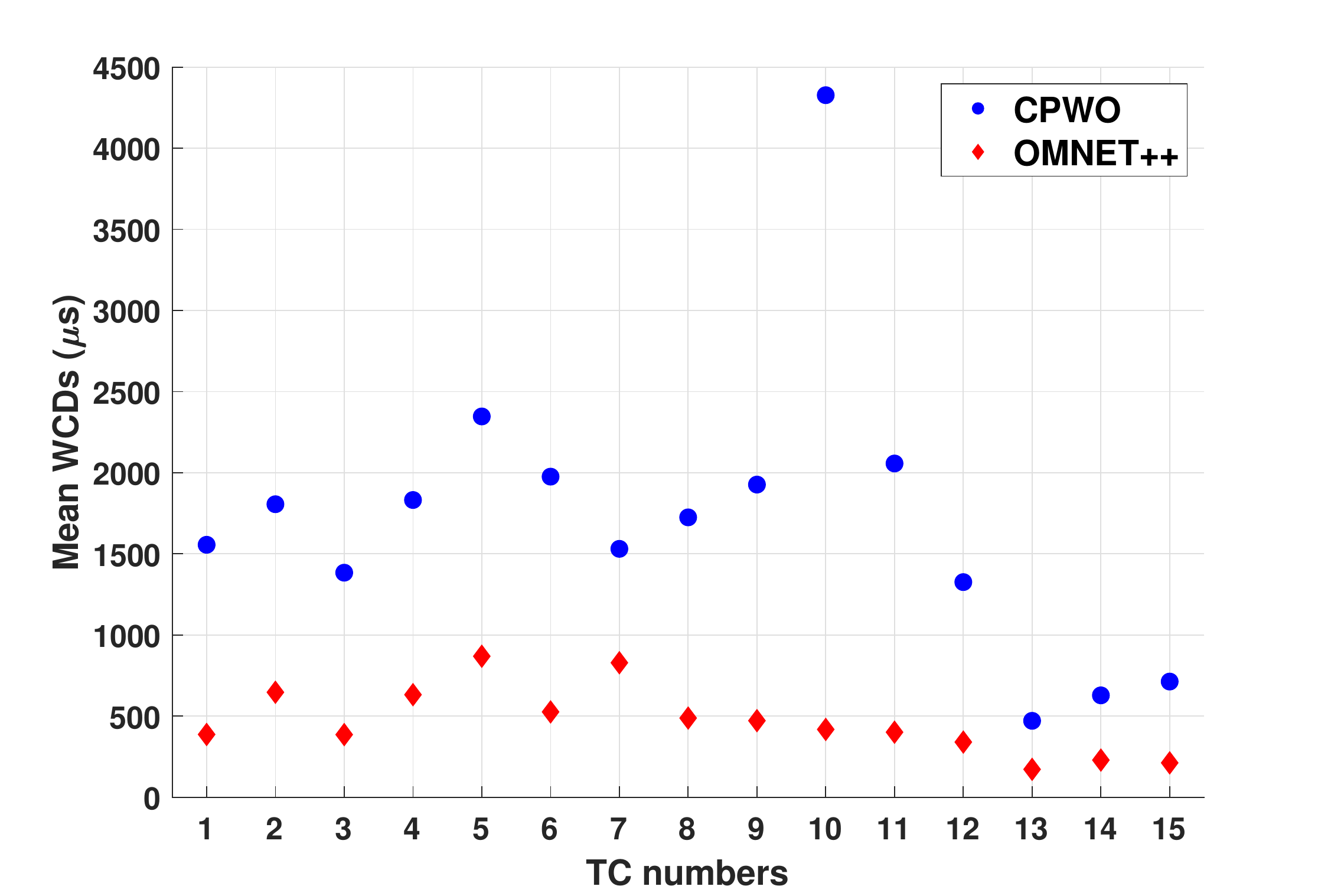}
         \caption{Mean WCDs vs. simulated delays for FWND}
         \label{fig:WCDComparison}
     \end{subfigure}
     \hfill
     \begin{subfigure}[b]{0.45\textwidth}
         \centering
         \includegraphics[width=\textwidth]{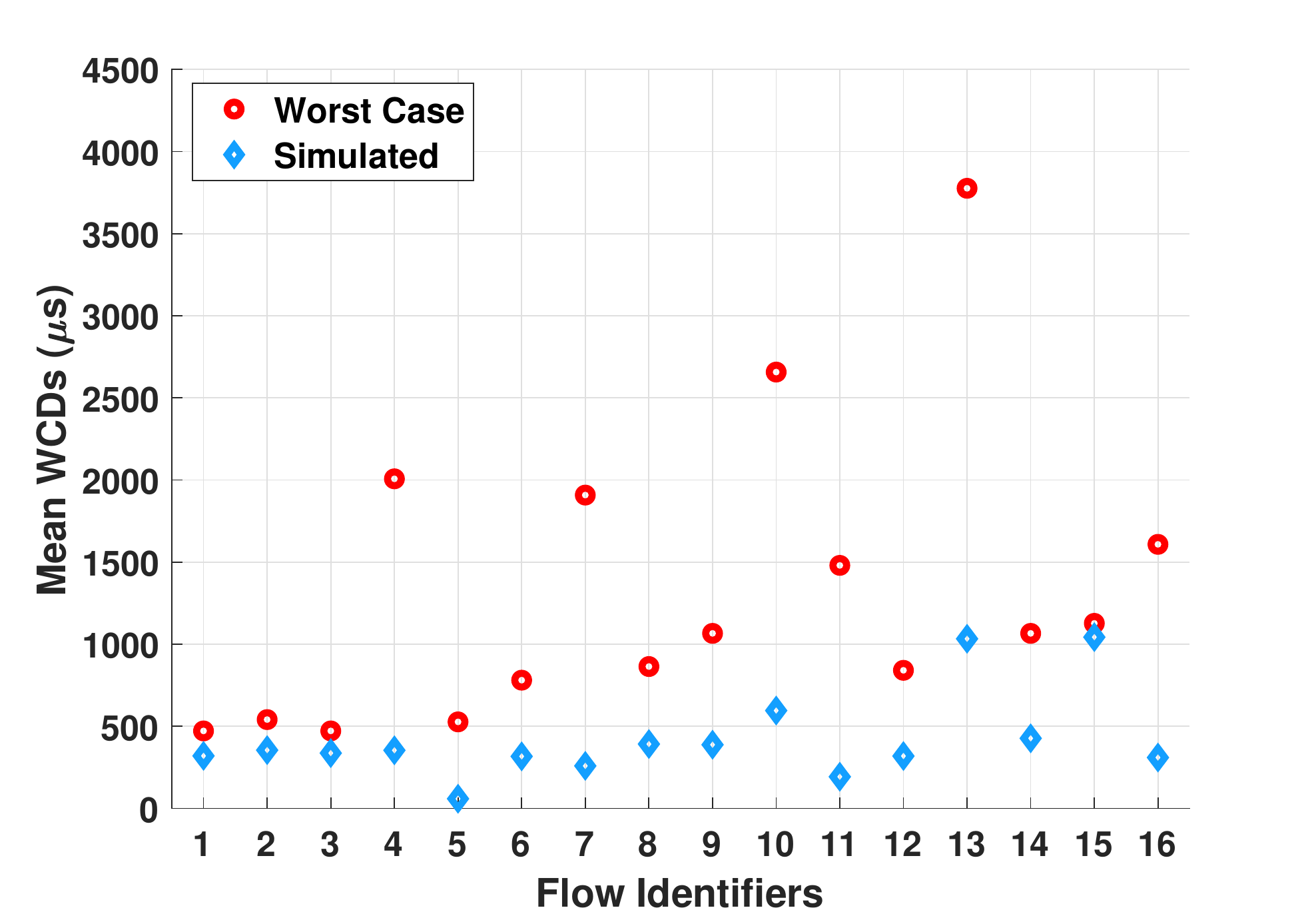}
         \caption{Mean WCDs comparison for TC12}
         \label{fig:TCWCD}
     \end{subfigure}
     \caption{Worst-case delay comparison with simulated response times.}
\end{figure}

We have used the OMNET++ simulator with the TSN  NeSTiNg extension~\cite{nesting_2019} to validate the generated GCLs. Thus, we have synthesized the GCLs for all approaches on all synthetic test cases, and we have observed that the GCLs are correct and the simulation behaves as expected. The mean WCDs of CPWO for the synthetic test cases and the worst-case latency observed during multiple OMNET++ simulations (with the windows from CPWO) are depicted in Fig.~\ref{fig:WCDComparison}. As expected, the latency values reported by OMNET++ are smaller than the WCDs, as reported by the WCD Analysis from~\cite{Zhao20:JIoT}. This is because a simulation cannot easily uncover the worst-case behavior. However, the simulation indicates the average behavior, and small delays mean that even for unscheduled/unsynchronized end systems, we are able to obtain solutions that are not only schedulable (WCDs are smaller than the deadlines) but also have good average behavior, where most of the time the delays are reasonable, even smaller than the static schedules obtained by \emph{0GCL} and \emph{FWND} for scheduled and synchronized ESs. The pessimism result of the WCD analysis is unavoidable in systems with un-synchronized and/or unscheduled end-systems; in practice, however,  simulated delays are much smaller, as can be seen in Fig.~\ref{fig:WCDComparison} and Fig.~\ref{fig:TCWCD}. We also show in Fig.~\ref{fig:TCWCD} the simulated delays and WCDs for all flows of TC12. All the flows are schedulable, and, as expected, the simulated delays are smaller than the WCDs, calculated with the worst-case delay analysis derived in the work from~\cite{Zhao20:JIoT}.

\subsection{Scalability Evaluation}
\label{sec:scalibility}

We have investigated the scalability of CPWO on 6 larger test cases (TC1 to TC6), that have up to 120 devices (75 ESs and 45 SWs) and 500 flows. The results and the details of the test cases are presented in Table~\ref{tab:scalabilityresults}, where columns 2, 3, and 4 show the number of flows, end-systems, and switches, respectively. Columns 5, 6, and 7 show the mean WCD of flows in $\mu$s, the largest deadline of all flows in $\mu$s, and the objective value~$\Omega$, related to bandwidth, see Eq.~(\ref{eq:avgbandwidth}). CPWO was able to generate schedulable solutions in all cases. Furthermore, CPWO is optimize the schedules for minimum bandwidth usage and has generated solutions that, besides being schedulable, have mean WCDs on average 14\% smaller than the respective deadlines in all test cases.

\begin{table}[t!]
\small
\caption{Scalability evaluation of CPWO}
\label{tab:scalabilityresults}
\centering
\begin{adjustbox}{width=0.99\textwidth}
\small
\begin{tabular}{l|c|c|c|c|c|c}
\hline
\textbf{No.}&\textbf{Total No.}&\textbf{Total No.}&\textbf{Total No.}&\textbf{Mean WCDs}&\textbf{Largest deadline}&\textbf{$\Omega$}\\
&\textbf{of Flows}&\textbf{of ESs}&\textbf{of SWs}&\textbf{$\mu$s}&\textbf{$\mu$s}&\textbf{($\times 1000$)}\\
\hline
\hline
TC1 & 100 &50 & 35& 3,226 & 4,000 & 249\\
TC2 & 150 &55 & 40& 3,521 & 4,000 & 366\\
TC3 & 200 &60 & 40& 4,387 & 5,000 & 396\\
TC4 & 300 &65 & 40& 4,911 & 6,000 & 468\\
TC5 & 400 &70 & 45& 5,210 & 6,000 & 498\\
TC6 & 500 &75 & 45& 4,399 & 5,000 & 511\\
\hline
\end{tabular}
\end{adjustbox}
\end{table}

\section{Conclusions}
\label{sec:Conclusion}
We have presented a novel, more flexible heuristic schedule synthesis approach for TSN networks, which decouples the frame scheduling from the generation of time-aware shaper (TAS) windows. Our method eliminates the often-unrealistic constraint that end systems are scheduled and synchronized (i.e., they have TSN capabilities) required by previous methods to provide real-time guarantees for critical traffic in IEEE 802.1Qbv TSN networks. Our approach intertwines an existing worst-case delay analysis method with a CP-solver into a novel and scalable heuristic approach that uses a Tabu Search metaheuristic search strategy in the CP-solver. Furthermore, to improve scalability, we have proposed a novel \emph{proxy function} which can be parametrized to trade-off runtime performance for search-space pruning in the CP-model. We evaluated our approach using synthetic and real-world test cases, comparing it with existing mechanisms and validated the generated schedules using OMNET++.

\bibliography{paper}

\end{document}